\documentclass[twocolumn,nopacs,floatfix,amsmath,nofootinbib,amssymb,preprintnumbers]{revtex4-2}
\usepackage{graphicx,color,dcolumn,booktabs,bm,siunitx,multirow,mathrsfs}
\usepackage{txfonts}
\usepackage{xcolor}
\usepackage{ulem}
\usepackage{slashed}
\usepackage[colorlinks,
            citecolor=blue,
            anchorcolor=red,
            menucolor=red,
            linkcolor=red,
            filecolor=red,
            runcolor=red,
            urlcolor=blue,
            frenchlinks=red]{hyperref}

\begin{document}

\title{Lattice QCD constraints on pion electroproduction off a nucleon}

\author{Yu Zhuge$^{1,2,3}$}
\author{Zhan-Wei Liu$^{1,2,3}$}\email{liuzhanwei@lzu.edu.cn}
\author{Derek~B.~Leinweber$^4$}\email{derek.leinweber@adelaide.edu.au}
\author{Anthony~W.~Thomas$^4$}\email{anthony.thomas@adelaide.edu.au}

\affiliation
{
    $^1$School of Physical Science and Technology, Lanzhou University, Lanzhou 730000, China\\
    $^2$Research Center for Hadron and CSR Physics, Lanzhou University and Institute of Modern Physics of CAS, Lanzhou 730000, China\\
    $^3$Lanzhou Center for Theoretical Physics, MoE Frontiers Science Center for Rare Isotopes, Key Laboratory of Quantum Theory and Applications of MoE, Key Laboratory of Theoretical Physics of Gansu Province, Gansu Provincial Research Center for Basic Disciplines of Quantum Physics, Lanzhou University, Lanzhou 730000, China\\
    $^4$CSSM, Department of Physics, Adelaide University, South Australia 5005, Australia
}

\begin{abstract}
    Very recently, a lattice QCD collaboration has explored threshold pion electroproduction near the physical pion mass and has simulated the relevant multipole amplitudes. Different multipole amplitudes are usually entangled in experimental data, and thus extracting each of them independently from first principles provides additional essential constraints on phenomenological theories. We use nonperturbative Hamiltonian theory to investigate the electroproduction process, providing an advanced approach with additional two-particle coupled channels to acquire the physical electric dipole amplitudes from the original lattice QCD data. We note that future lattice QCD simulations of the electric dipole amplitudes at higher energies will be much closer to their physical counterparts than the current ones near threshold. In addition, we obtain a new expression which, like that of Lellouch–L{\"u}scher, depends only on the final-state interactions but provides both the real and imaginary parts of the transition amplitudes.
\end{abstract}

\maketitle


    {\it Introduction} --- Deep inelastic scattering reveals the parton structure of nucleons at large energy scales where perturbative QCD is very successful. At low energies, electroproduction off the nucleon exhibits the inherently non-perturbative dynamics of QCD, which play a central role in mediating the strong forces that bind matter. 
    
    Studies of pion photo- and electroproduction have a long and rich history. Dispersion relations were first employed to analyze this process in the 1950s by Chew, et al.~\cite{Chew:1957tf}. In the mid to late 20th century, low-energy theorems provided a remarkably accurate description of charged-pion photoproduction~\cite{Kroll:1953vq, Walker:1963zzb, DeBaenst:1970dqx, Vainshtein:1972ih, Rossi:1973wf, Salomon:1983xn, Davidson:1988yz, Drechsel:1992pn}. Later, chiral perturbation theory (ChPT) has been extensively employed to analyze pion photo- and electroproduction across all charge channels~\cite{Bernard:1991rt, Bernard:1993bq, Mai:2012wy, Hilt:2013uf, GuerreroNavarro:2019fqb, GuerreroNavarro:2020kwb, Rijneveen:2021bfw, Kang:2024fsf}. Meanwhile, the wealth of meson electroproduction data gathered at various facilities has allowed further development of these low-energy effective theories in the non-perturbative region of QCD~\cite{Asner:2008nq, CLAS:2009ces, A1:2017fvu, Beck:2017wkb, LEPS:2017pzl, Ohnishi:2019cif, BGO-OD:2019utx, Burkert:2025coj, Doring:2025sgb}. With the availability of extensive experimental data, partial-wave analyses of the photoproduction process have been performed by the ANL-Osaka~\cite{Matsuyama:2006rp}, SAID~\cite{Briscoe:2023gmb}, MAID~\cite{Drechsel:2007if}, Jülich~\cite{Ronchen:2014cna} and other groups. In addition, the chiral quark model has also been applied to analyze the experimental data and explore the properties of the relevant resonances~\cite{Li:1994cy, Xiao:2015gra, Zhao:2002id}. Pion electroproduction near threshold not only helps explore the electromagnetic properties of nucleons but also improves our understanding of the background in the higher energy region where the hadron resonances can be studied. 
    
    Different multipole amplitudes are entangled in experiments, while lattice QCD can simulate them independently from the first principles of QCD. Their combination will make the mechanism of pion electroproduction even clearer. Very recently, the multipole amplitudes near threshold for pion electroproduction --- evaluated as functions of the spacelike photon momentum transfer --- were extracted from lattice QCD~\cite{Gao:2025loz}. To improve the precision in extracting nucleon matrix elements, it is crucial to suppress excited-state contamination, for example, by employing $\pi N$ interpolating operators~\cite{Barca:2022uhi, Alexandrou:2024tin, Gao:2025loz}. By solving the generalized eigenvalue problem and constructing excited-state operators, the excited-state contamination has been assessed for pion electroproduction. Reducing this contamination is crucial~\cite{Gao:2025loz}.
    
    Nonperturbative Hamiltonian theory (NPHT) provides a powerful framework for relating finite-volume spectra from lattice QCD to infinite-volume scattering observables~\cite{Liu:2015ktc, Abell:2023nex}.  In Refs.~\cite{Guo:2022hud,Zhuge:2024iuw}, we investigated  pion photoproduction, extracting the relevant multipole amplitudes and comparing our results with those of established partial-wave analyses. 

    In this Letter we take advantage of the very recent lattice data to extend our previous NPHT-based analysis of pion photoproduction to the case of pion electroproduction. This process, involving a virtual photon, is sensitive to the internal structure of the nucleon and presents additional theoretical challenges because of the $q^2$-dependence of the amplitudes. Our calculations extend previous finite-volume corrrection analyses by including the $\eta N$ and $K \Lambda$ coupled channels and showcasing their role in precision determinations of the physical observables. After analyzing the real electric dipole amplitudes on the lattice, we use NPHT to extract both the real and the imaginary parts of the physical amplitudes in infinite volume. 
    
    Since the first excited $\pi N$ contribution has also been included when dealing with contamination in the lattice QCD work ~\cite{Gao:2025loz}, the pion electroproduction at energies larger than the $\pi N$ threshold could be simulated soon. Our results show that the finite-volume effects on the electric dipole amplitudes are much smaller at the excited eigenenergies than at threshold. Therefore, such lattice QCD simulations will provide more powerful constraints on electroproduction.   



{\it The electric dipole amplitudes in the infinite volume} --- We outline the main framework of pion electroproduction in infinite volume. Since the electron–electron interaction is purely electromagnetic in the pion electroproduction process $ e + N \rightarrow e + \pi + N$,  the hadronic subprocess
\begin{align} 
    \gamma^*(q) + N(p) \rightarrow \pi(k) + N(p^\prime) 
\end{align} 
can be treated separately~\cite{Hilt:2013fda, Cao:2021kvs}. The T matrix $T_{\pi N,\gamma^* N}^{\lambda_{\gamma^*},\lambda_{N}}$ can be split into two parts, the tree level transition potential $\gamma^* N\overset{\rm EM}{\rightarrow}\pi N$ and the final-state interaction (FSI) corrections $\gamma^* N\overset{\rm EM}{\rightarrow}\pi N/\eta N/...\overset{\rm FSI}{\rightarrow}\pi N$, 
\begin{align}
\label{finalT}
	&T_{\pi N,\gamma^* N}^{\lambda_{\gamma^*},\lambda_{N}}(k,q;E_{\rm cm})=V_{\pi N,\gamma^* N}^{JLS;\lambda_{\gamma^*},\lambda_{N}}(k,q)+\sum_{\alpha}\int\mathrm{d}k'k^{\prime2}\notag\\ &\qquad\times
 V_{\alpha,\gamma^* N}^{JLS;\lambda_{\gamma^*},\lambda_{N}}(k',q)
	\frac{1}{E_{\rm cm}-\omega_{\alpha}(k')+i\epsilon}T_{\pi N,\alpha}(k,k';E_{\rm cm}) \, ,
\end{align}
where $\alpha$ represents the involved coupled channels $\pi N/\eta N/...$ and $\omega_{\alpha}$ is its energy. The superscripts are related to the angular momentum. See Ref.~\cite{Zhuge:2024iuw} for further details. The tree level amplitude $V_{\alpha,\gamma^* N}^{JLS;\lambda_{\gamma^*},\lambda_{N}}$ can be obtained from the $s$-, $u$-, $t$-, and contact Feynman diagrams with relevant effective Lagrangians~\cite{Zhuge:2024iuw}. The strong-interaction scattering amplitudes $T_{\pi N,\alpha}$ can be obtained with NPHT which is successful in describing both the relevant experimental scattering data and the resonance mass spectra from lattice QCD simulations~\cite{Liu:2015ktc, Abell:2023nex}. The $S_{11}$ partial wave amplitude related to the $E_{0+}$ is
\begin{align}
    E_{0+}(E_{\rm cm}, q^2)=\frac{\pi \, m_N\sqrt{\omega_{\pi}(k_{\rm on})\, |q_{0}|}}{4E_{\rm cm}}T^{\lambda_{\gamma^*}=1,\lambda_N=1/2}_{\pi N,\gamma^* N}(k,q;E_{\rm cm}) \, .
\end{align}

For the off-shell photon, we introduce nucleon and pion electromagnetic form factors at the $\gamma^*NN$ and $\gamma^*\pi\pi$ vertices, respectively, to describe the $q^2$ dependence more in a comprehensive manner. The nucleon electromagnetic form factors take the dipole form which can describe the small $q^2$ region well~\cite{Thomas:2001kw, Perdrisat:2006hj, Alberico:2008sz}
\begin{align}\label{GEp}
    G^p_{E}(q^2)=\frac{G^p_{M}(q^2)}{\mu_{p}}=\left(1-\frac{q^2}{0.71\,{\rm GeV^2}}\right)^{-2} \, ,
\end{align}
\begin{align}\label{GEn}
    G^n_{E}(q^2)\frac{1-B\tau}{-A\tau}=\frac{G^n_{M}(q^2)}{\mu_{n}}=\left(1-\frac{q^2}{0.71\,{\rm GeV^2}}\right)^{-2} \, ,
\end{align}
where $\tau=\frac{q^2}{4m^2_{N}}$, $A = 1.68$ and $B=3.63$~\cite{Alberico:2008sz}. The Dirac and Pauli form factors $F_1(q^2)$ and $F_2(q^2)$, which we employ, are related to $G_E(q^2)$ and $G_M(q^2)$~\cite{Thomas:2001kw} in the standard manner. The pion form factor is parameterized as a monopole form~\cite{NA7:1986vav}
\begin{align}\label{Fpi}
    F_{\pi}(q^2)=\left(1-\frac{q^2 \, \langle r^2_{\pi}\rangle}{6}\right)^{-1} \, ,
\end{align}
where $\langle r^2_{\pi}\rangle = 0.431$ $\rm fm^2$. The pion form factor should also be included in the original contact term to keep the gauge invariance, provided that $F_\pi(q^2)=F_1^p(q^2)-F_1^n(q^2)$~\cite{Nozawa:1989gy, Drechsel:1998hk}. Indeed, this condition is satisfied well at small $q^2$. 
\begin{figure}[tbp]  
	\centering
	\begin{tabular}{c}
		\includegraphics[width=8.5cm]{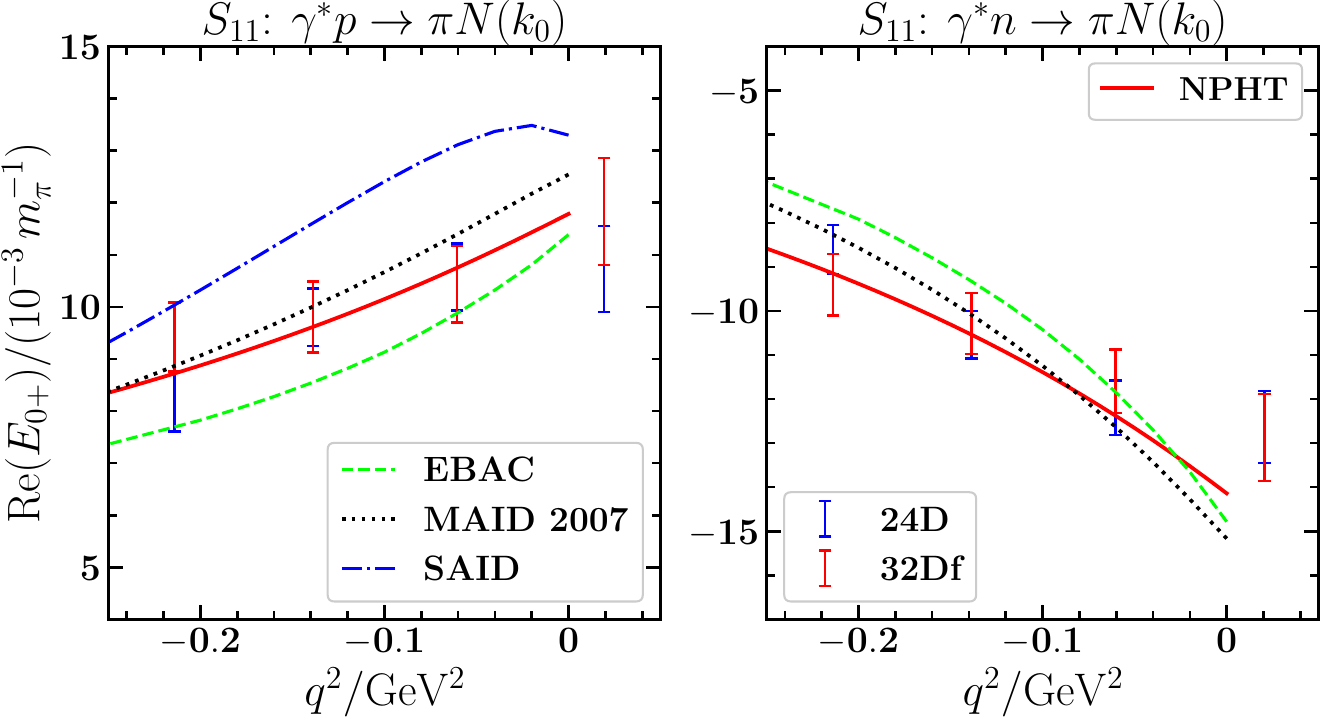}
	\end{tabular}
	\caption{The multipole amplitude $E_{0+}$ (in units of $10^{-3}/m_{\pi}$) at the $\pi N$ threshold, with isospin $I_{\pi N}=1/2$. Our results are shown as the red solid lines. Other line styles represent different partial-wave analysis groups: SAID (blue dash-dotted)~\cite{GWU, Briscoe:2023gmb}, EBAC (green dashed)~\cite{site:SL}, and MAID 2007 (black dotted)~\cite{maid, Drechsel:2007if}. The data points are from lattice QCD simulations with the spatial extent $L=4.6$ fm and 24D or 32Df ensembles~\cite{Gao:2025loz}, after application of the Lellouch-L\"uscher factor.}
\label{basestate}
\end{figure}

Since the pion mass used in the lattice QCD calculation is about 143 MeV~\cite{Gao:2025loz}, very close to the physical value, we employ the physical hadron masses throughout our analysis. The infinite-volume amplitude $E_{0+}$ at the $\pi N$ threshold is shown in Fig.~\ref{basestate} for the transition $\gamma^*N\rightarrow\pi N(k_{0})$. Here, the lattice QCD results have been brought to infinite volume through the application of the Lellouch-L\"uscher factor.

Since near the $\pi N$ threshold, the $\eta N$ and $K \Lambda$ channels remain closed, the FSI effects encoded in the second term of Eq.~(\ref{finalT}) are weak. Instead, the dominant contribution comes from the direct transition amplitude, $V_{\pi N,\gamma^* N}^{JLS;\lambda_{\gamma^*},\lambda_{N}}(k,q)$, in which the contact term alone accounts for over 85\% of the strength. As a consequence of gauge symmetry, the contact term is proportional to the pion-nucleon coupling, $f_{\pi NN}$, which is usually quoted at the pion pole~\cite{Ericson:1988gk, Ericson:2000md, Reinert:2017usi, Reinert:2020mcu}. The recent determination in Ref.~\cite{Reinert:2020mcu} provides the best empirical value, $f_{\pi NN} = 0.984\pm 0.007$. We employ a Gaussian form factor $u_{\pi NN} = \exp\left(-\frac{\vec{k}^2+m^2_{\pi}}{2\Lambda_{\pi NN}^2}\right)$, with $\Lambda_{\pi NN}=450$ MeV~\cite{Reinert:2017usi,Reinert:2020mcu}. All of the other parameters used here are the same as those determined in our earlier study~\cite{Zhuge:2024iuw}, with no additional tuning. In Fig.~\ref{basestate}, we see that our predictions, given by the red solid lines, describe the lattice QCD results well, within $1\sigma$. We also plot other partial-wave analysis results in the figure and it is clear that there are some obvious discrepancies between these different models. Some of these analyses covered a very large energy region rather than just the threshold region. However, they still have similar trends as $q^2$ varies. 



{\it The electric dipole amplitudes in finite volume and extrapolations} --- L{\"u}scher’s formula provides a powerful framework for relating finite-volume spectra from lattice QCD to infinite-volume experimental observables~\cite{Luscher:1985dn, Luscher:1986pf, Luscher:1990ux}. This formalism is extended to connect finite-volume matrix elements with infinite-volume weak decay amplitudes~\cite{Lellouch:2000pv}. The $\pi N$ rescattering effects were taken into account, using this method, in Ref.~\cite{Gao:2025loz}, where the so-called Lellouch–L{\"u}scher factor, relating finite- and infinite-volume states, was implemented  using the relevant $\pi N$ scattering length. 

In lattice QCD calculations, the extraction of multipole amplitudes begins with the computation of four-point correlation functions, $C_{N\pi J N}$ \cite{Gao:2025loz}. After applying the Lellouch-L{\"u}scher factor, the physical amplitudes in infinite volume are obtained. These may then be compared with predictions from other theoretical models~\cite{Gao:2025loz}.

In this Letter, we start from the original electric dipole amplitudes $E_{0+}^L$ on the lattice and provide another method to obtain the physical amplitudes $E_{0+}$ via NPHT. Not only are the real parts equivalent to those given by the Lellouch-L{\"u}scher method, but we can also obtain the imaginary parts. Moreover, additional relevant two-particle scattering channels are easily incorporated into the analysis. The results may then be compared to those obtained using ChPT and other phenomenological approaches. 

In a cubic lattice of spatial extent $L$, the momentum is quantized as
\begin{align}
    \vec{k}_{\vec{n}} = \frac{2\pi}{L}\vec{n}, \quad
    k_n \equiv |\vec{k}_{\vec{n}}| , \quad \vec{n}=(n_x,n_y,n_z) \in \mathbb{Z}^3 \, .
\end{align}
In a lattice QCD simulation, momentum-projected states evolve and ultimately form genuine finite-volume eigenstates $|G(n)\rangle$. The ground eigenstate, $|G(0)\rangle$, consists of 99.8\% $|\pi N(k_0)\rangle$ when $L=4.6$ fm. We compute the transition amplitude, $\gamma^* N \rightarrow G(n)$, in the finite volume, which corresponds to the amplitude from the original lattice simulations prior to the application of the Lellouch–L{\"u}scher factor. The $1/2^-$ eigenvector $|G(n)\rangle$ in a box can be expanded in terms of the basis states as~\cite{Abell:2023nex}
\begin{align} \label{expandEN}
    |G(n)\rangle = &c_1|N_1\rangle + c_2|N_2\rangle \notag\\
    &+\sum_{n_1, n_2}\left[c^{(1)}_{n_1}|\pi N(k_{n_1})\rangle + c^{(2)}_{n_2}|\eta N(k_{n_2})\rangle+ ...\rangle\right] \, ,
\end{align}
where $|N_1\rangle$ and $|N_2\rangle$ are the bare states associated with the two low-lying $S_{11}$ nucleon resonances. We define the amplitude $E_{0+}^L$ in the finite volume, corresponding to the transition $\gamma^* N \rightarrow G(n)$, as follows
\begin{align} \label{Etilde}
    E_{0+}^L =&\frac{\pi \, m_N\sqrt{\omega_{\pi}(k_{\rm on})\, |q_{0}|}}{4E_{G(n)}} \left\{\sum_{i} c_i \sqrt{\frac{4\pi}{C_3(n)}} \left( \frac{L}{2\pi} \right)^{\frac{3}{2}} \, G_{N_i,\gamma^*N}  \right. \notag\\
    & \left.+\sum_{\alpha,n_i} c^{(i)}_{n_i} \sqrt{\frac{C_3(n_i)}{C_3(n)}} V_{\alpha(n_i),\gamma*N} \right\} \, ,
\end{align}
where the degeneracy $C_3(n)$ denotes the number of integer triplets $(n_x, n_y, n_z)$ satisfying $n = n^2_x + n^2_y + n^2_z$ for each $n$. $G_{N_i,\gamma^*N}$ represents the coupling between the bare $N^*$ and $\gamma^*N$~\cite{Zhuge:2024iuw} and $V_{\alpha,\gamma*N}$ is the infinite-volume tree level transition potential; except that we need to subtract the $s$-channel bare-state exchange contributions, as these have been included in $G_{N_i,\gamma^*N}$.

The nucleon and pion electromagnetic form factors have also been calculated in lattice QCD, and they are consistent with the parameterizations obtained from direct experimental measurements within the $q^2$ region considered in this work~\cite{CSSM:2014knt, Shanahan:2014cga, Shanahan:2016iue, Wang:2020nbf, Gao:2021xsm}. Therefore, we adopt the parameterized forms given in Eqs.~(\ref{GEp})-(\ref{Fpi}), for both the infinite-volume and finite-volume cases.

Now we allow the coupling constant $f_{\pi NN}$ to vary, to fit the original lattice QCD data with $E_{0+}^L$ in the isospin $I_{\pi N}=\frac{1}{2}$ channel. This yields the best-fit value 
$f_{\pi NN} = 0.96\pm0.05$ with $\chi^2_{\rm d.o.f}=0.43$, 
which is completely consistent with the empirical value cited earlier~\cite{Reinert:2020mcu}.
These results indicate that our approach provides a reasonable description of the near-threshold electroproduction process.

One can also extrapolate the physical amplitudes $E_{0+}$ obtained from the lattice QCD data using NPHT. Our extrapolated results are shown as red solid lines in Fig.~\ref{basestate}, alongside the data points obtained using the Lellouch–L{\"u}scher method. As seen in the figure, the two results are very similar. In our approach, the ratio of infinite- to finite-volume amplitudes is approximately 1.11$\sim$1.13, while the finite-volume correction factor extracted using the Lellouch–L{\"u}scher method in Ref.~\cite{Gao:2025loz} is around 1.15. Both indicate that rescattering effects near the $\pi N$ threshold are not very large. The slight deviation can be attributed to the fact that the finite-volume energy $E_{G(0)}$ is slightly below the $\pi N$ threshold $E_{\pi N(0)}$, whereas the infinite-volume $E_{0+}$ should be measured at $E_{\pi N(0)}$ rather than $E_{G(0)}$.

Lattice QCD has also made remarkable progress in studying the hadron spectrum typically with the focus on the masses of the resonances. Moreover, through the well-known L{\"u}scher formalism, one can relate the finite-volume energy levels obtained from lattice simulations to experimental scattering data such as the phase shifts of the relevant reactions. This connection extends the extrapolation to the widths of the resonances~\cite{Hansen:2019nir}. Alternatively, NPHT can also study the phase shifts associated with lattice QCD observations and the channel thresholds in the finite volume. This  then allows us to calculate the poles in the infinite volume, with real and imaginary parts that characterize the resonances~\cite{Yu:2023xxf, Wang:2025hew, Han:2025gkp}. Looking forward, extracting the imaginary parts of the electric dipole amplitudes $E_{0+}$ may offer further opportunities to study the electromagnetic properties of hadrons and give insight into their structure. 

The imaginary parts, ${\rm Im}E_{0+}$, come from the FSI effects and thus have the advantage of revealing this mechanism more clearly. They are studied widely and here we take ${\rm Im}E^{\gamma p\rightarrow\pi^0p}_{0+}$ near threshold as an example. It is proportional to the center-of-mass momentum $|\vec{k}_{\pi^+}|$ of the $\pi^+n$ system near threshold~\cite{Fernandez-Ramirez:2012vlt, Hilt:2013uf}:
\begin{align}\label{dirIm}
    {\rm Im}E^{\gamma p\rightarrow\pi^0p}_{0+}=\beta\frac{|\vec{k}_{\pi^+}|}{m_{\pi^+}} \, .
\end{align}
More precisely, the following expression was used in Ref.~\cite{Hilt:2013uf}
\begin{align}\label{beta2}
     {\rm Im}E^{\gamma p\rightarrow\pi^0p}_{0+}=\frac{|\vec{k}_{\pi^+}|}{m_{\pi^+}}\left(  \beta + \gamma \frac{q^{\rm lab}_{\gamma}-q^{\rm lab,thr}_{\gamma}}{m_{\pi^+}} \right) \, ,
\end{align}
where $q^{\rm lab}_{\gamma}$ is the photon momentum in the laboratory frame and $q^{\rm lab,thr}_{\gamma}$ corresponds to the value at the $\pi^+ n$ threshold. Our formalism naturally yields the imaginary parts ${\rm Im}E_{0+}$ in infinite volume after studying $E_{0+}^L$ on the lattice. The extracted values of $\beta$ and $\gamma$ are listed in Table~\ref{beta}. Comparing with other approaches in the table, one sees that the acquisition of the imaginary parts of the multipole amplitudes from the raw lattice QCD datasets is feasible. These results indicate that our approach provides a reasonable description of the near-threshold electroproduction process.
\begin{table}[tbp]
    \caption{The extracted $\beta$ and $\gamma$ of the $\mathrm{Im}E^{\gamma p\rightarrow\pi^0p}_{0+}$ based on lattice QCD simulations analysed with NPHT are compared with other models in units of $10^{-3} m_{\pi^+}$.}
    \centering
    \label{beta}
    \def\arraystretch{1.2}
    \begin{tabular}{c r S[table-format=-1.4] S[table-format=-1.4] S[table-format=-1.4]}
        \toprule
        & \multicolumn{1}{c}{\multirow{2}{*}{This work}}
        & \multicolumn{2}{c}{Ref.~\cite{Hilt:2013uf}} 
        & \multicolumn{1}{c}{\multirow{2}{*}{Ref.~\cite{Bernard:1995cj}}} \\
         \cmidrule(r){3-4}
        & 
        & {HBChPT} 
        & {RChPT} 
        & \\
        \hline
        $\beta$  & $2.97\pm0.16$  & 2.83 & 3.16 & 2.89 \\
        $\gamma$  &  $-1.71\pm0.40$  & -1.97 & -1.08 & \\
        \bottomrule
    \end{tabular}
\end{table}
%


{\it Reduction of finite-volume effects at higher energies} --- Usually the physics in the finite and infinite volumes is very different. For example, in the finite volume one finds eigenstates of QCD, whereas in the real world one encounters resonances and two-hadron scattering states. Frequently, the lattice QCD mass spectrum would be transformed into the phase shifts in the physical world through the L\"uscher formula, and such transformation can be extremely complicated when a few coupled channels are involved. If an observable suffers small finite-volume effects, that is it is not modified much but rather has direct numerical correspondence with the infinite volume value, it is very attractive for lattice QCD simulations.

\begin{figure}[tbp]
	\centering
	\begin{tabular}{c}
        \includegraphics[width=8cm]{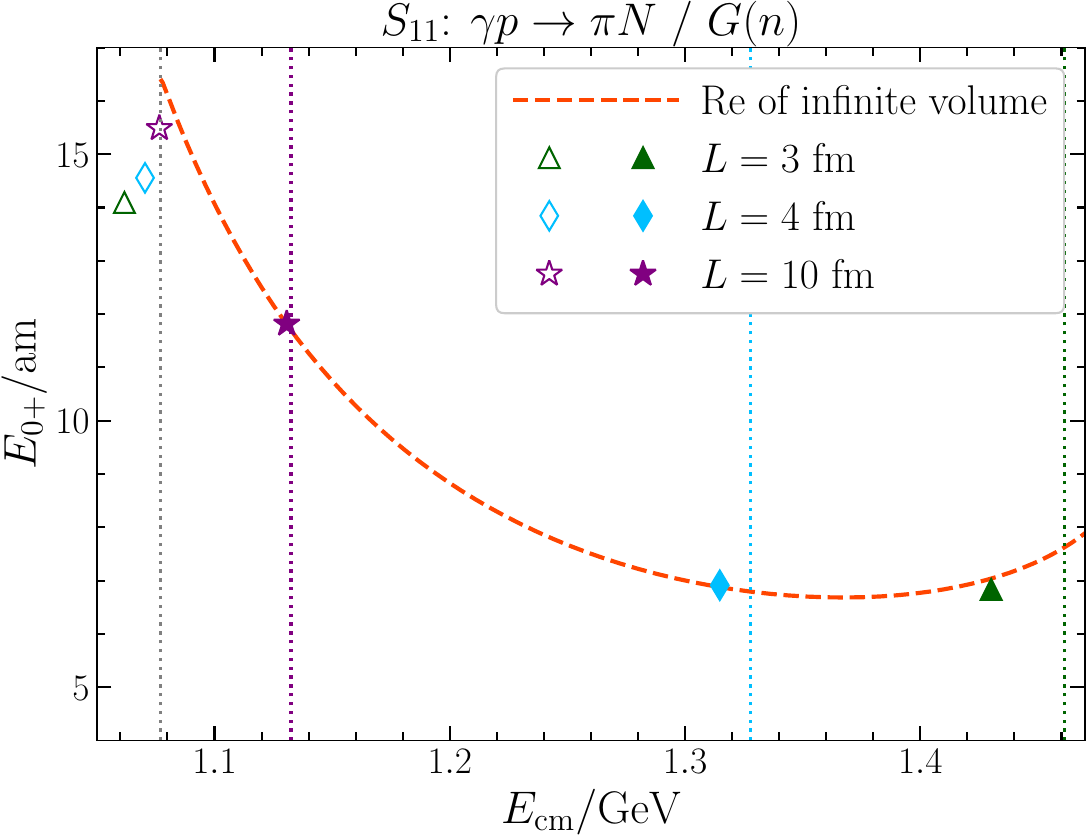}
	\end{tabular}
	\caption{The dependence of the multipole amplitude $E_{0+}$ on the center-of-mass energy $E_{\rm cm}$ for $q^2 = 0$. The red dashed line represents the infinite volume $E_{0+}$, and the hollow and filled data points refer to the finite-volume $E^L_{0+}$ corresponding to the two lowest finite-volume eigenstates $|G(0)\rangle$ and $|G(1)\rangle$, respectively. The gray vertical line marks the $|\pi N(k_0)\rangle$ threshold. The green, blue, and purple vertical lines indicate the noninteracting $|\pi N(k_1)\rangle$ levels with different box sizes $L$. }
	\label{fitE}
\end{figure}

We have seen that the difference between $E_{0+}$ and $E_{0+}^{\rm 4.6~fm}$ is not very large near the $\pi N$ threshold. In Fig.~\ref{fitE} we plot three values of $E_{0+}^{L}$ with hollow symbols, for volumes corresponding to $L=$ 3, 4, and 10 fm obtained from our NPHT analysis. This illustrates clearly that $E_{0+}^{L}$ with larger $L$ goes closer to the real part of $E_{0+}$ in the infinite volume near the $\pi N$ threshold. It is not a trivial check if one notices that the procedures to obtain the $E_{0+}$ and $E_{0+}^{L}$ are so different, especially when dealing with FSI. 

In addition to the ground state $G(0)$, we also explore the electric dipole amplitudes $E_{0+}^{L}$ for the first excited state $G(1)$. We show three $E_{0+}^L$ for $G(1)$ with filled shapes in Fig.~\ref{fitE}. NPHT provides that the eigenstates $G(1)$ are dominated by $\pi N(k_1)$ and thus in Fig.~\ref{fitE} the filled points are close to noninteracting $\pi N(k_1)$ levels, respectively. Obviously the finite volume effects of $E_{0+}$ are much smaller for the first excited states $G(1)$ compared to those for the ground states $G(0)$, even though the $G(1)$ points deviate from the corresponding vertical dotted lines more than those for $G(0)$.    

Currently, only the amplitudes for $\gamma^* N \rightarrow G(0)$ have been simulated on the lattice~\cite{Gao:2025loz}. With interpolating operators beyond the traditional three-quark ones, the mass spectrum of excited states has been determined with small uncertainties in the near-physical pion-mass region. Examples include the $1/2^-$ nucleon excited states calculated by the Coordinated Lattice Simulations consortium~\cite{Bulava:2022vpq} and the $1/2^-$ $\Lambda$ excited states studied by the Baryon Scattering collaboration~\cite{BaryonScatteringBaSc:2023ori, BaryonScatteringBaSc:2023zvt} and these lattice QCD results were well described by NPHT~\cite{Abell:2023nex, Liu:2023xvy}. Moreover, the $\pi N(k_1)$ interpolating operator has been included to decrease the contamination in the lattice QCD simulation of $\gamma^* N \rightarrow G(0)$. Therefore, it seems both timely and very worthwhile to explore the electromagnetic properties of the highly excited states further on the lattice.


{\it Improvement of the Lellouch–L{\"u}scher formula} --- The reduction of finite-volume effects in $E_{0+}/E^L_{0+}$ can also be demonstrated using the Lellouch–L{\"u}scher formula~\cite{Lellouch:2000pv, Gao:2025loz} as
\begin{align} \label{eq_fLL}
    F_{\rm LL}=\left|\frac{E_{0+}|_{E_{G(n)}}}{E_{0+}^{L}|_{E_{G(n)}}}\right| = \sqrt{\frac{2\pi \, C_3(n)}{(kL)^3} \left( {\tilde k}\frac{d\phi}{d{\tilde k}} + k \frac{d\delta}{dk} \right)} \, ,
\end{align}
where ${\tilde k} \equiv \frac{kL}{2\pi}$, the function $\phi({\tilde k})$ is related to the $\zeta$ function~\cite{Lellouch:2000pv} and $\delta(k)$ is the phase shift. The comparison for $E_{0+}/{E_{0+}^{L}}$ of the state $G(1)$ between the NPHT and the Lellouch–L{\"u}scher formula is listed in Table~\ref{fLL}, showing very close numerical agreement. This confirms, consistently with Fig.~\ref{fitE}, that finite-volume effects decrease for the first excited state $G(1)$.

As is well known, the Lellouch–L{\"u}scher formula can only extrapolate the absolute value $|E_{0+}|$, and thus another extrapolation method with which the real and imaginary parts can both be obtained is needed. If separable potentials of the form $V_{\alpha,\alpha'}(k,k')=h_\alpha(k) \, h_{\alpha'}(k')$ are employed~\cite{Veit:1984jr} where $\alpha^{(\prime)}=\gamma N, \pi N, \eta N,... $, we find another factor which also depends only on the FSIs, as does the Lellouch–L{\"u}scher formula
\begin{align}\label{eq:Fsep}
    F_{\rm sep} = \frac{E_{0+}|_{E_{G(1)}}}{E_{0+}^{L}|_{E_{G(1)}}} = \frac{2\sqrt{6}\pi^2 \, T_{\pi N,\pi N}\left(E_{G(1)}, k_{G(1)}, k_1\right)}{L^3 U^{\rm bind}_{1}c^{\pi N}_{1}} \, ,
\end{align}
where $c^{\pi N}_{1}=\langle \pi N(k_1)|G(1) \rangle =1+O(L^{-6})\approx 1$ is related to the $\pi N(k_1)$ component in the state $G(1)$, $U_1^{\rm bind} = E_{\pi N(1)}-E_{G(1)}$ denotes its finite-volume binding energy and $T_{\pi N,\pi N}$ is corresponding T matrix in infinite volume. The final expression for $F_{\rm sep}$ does not rely on the details in the separable potentials and may be still appropriate even if more complicated potentials are used. For example, Table~\ref{fLL} shows the consistency between the separable-potential factor $F_{\rm sep}$ and that obtained using NPHT. 
\begin{table}[tbp]
	\caption{The finite-volume correction factors $E_{0+}/E^L_{0+}$ with the NPHT, Lellouch-L{\"u}scher formula~\cite{Lellouch:2000pv, Gao:2025loz}, and separable potential formula of Eq.~(\ref{eq:Fsep}) for $\gamma p\to G(1)$.}
	\centering
	\label{fLL}
        \def\arraystretch{1.2}
    \begin{tabular}{c  S[table-format=-1.4] S[table-format=-1.4] | S[table-format=-1.4] S[table-format=-1.4] | S[table-format=-1.4] S[table-format=-1.4]}
        \toprule
        {$L/$fm}
		  & {${\rm Re}F_{\rm NPHT}$} & {${\rm Re}F_{\rm sep}$} & {${\rm Im}F_{\rm NPHT}$} & {${\rm Im}F_{\rm sep}$} & {$|F_{\rm NPHT}|$} & {$F_{\rm LL}$}\\
		\hline
        3          & 1.029  &  1.028  & 0.376 & 0.375  & 1.096   & 1.095  \\
        4          & 0.995  &  0.990  & 0.222 & 0.221  & 1.019 & 1.016 \\
        10         & 0.999  & 0.999 & 0.122 & 0.122 & 1.007 & 1.006\\
        \bottomrule
	\end{tabular}
 
\end{table}

The infinite-volume $E_{0+}$ should be measured at the $\pi N$ threshold $E_{\pi N(0)}$, rather than at the smaller finite-volume eigenenergy $E_{G(0)}$. These two energies differ by the binding energy $U_0^{\rm bind}$, which cannot be neglected for small $L$. For example, $U_0^{\rm bind}|_{3 \rm fm} = 15$ MeV. Therefore, the factor $E_{0+}|_{E_{\pi N(0)}}\,/\,E_{0+}^{L}|_{E_{G(0)}}$ requires a careful treatment. Within the NPHT framework, we have verified that this factor depends on both the electromagnetic potential and the FSIs. For example, the full NPHT calculation yields $F_{\rm NPHT}|_{3 \rm fm} = 1.164$. However, if only the electromagnetic potential with the contact diagram in the $\pi N$ channel were retained, the factor would change to 1.209. Consequently, $F_{\rm sep}$ and $F_{\rm LL}$, which are only dependent on FSIs, should be regarded as approximate methods in this case, with differences between them arising at higher orders. At the threshold, $F_{\rm sep}$ can be approximated as   
\begin{align} \label{eq_sep}
    F_{\rm sep} \approx \frac{2\pi a_{\pi N}}{\mu_{\pi N} U_0^{\rm bind} L^3 } 
\end{align}
where $a_{\pi N}$ is the $\pi N$ scattering length and $\mu_{\pi N}$ is the $\pi N$ reduced mass. It coincides with the Lellouch–L\"uscher result at order $O(L^{-1})$,
\begin{align} \label{eq_sep}
    F_{\rm sep/LL}=1 + \frac{Z_{00}(1,0)}{\pi}\frac{a_{\pi N}}{L} + O(L^{-2}) \, ,
\end{align}
where $Z_{00}(1,0)$ is the $\zeta$ function~\cite{Lellouch:2000pv}. In a 3 fm box, we obtain $F_{\rm sep}|_{3 \rm fm} = 1.145$ while $F_{\rm LL}|_{3 \rm fm} = 1.206$. These two approximations provide estimates at roughly the  $O(L^{-1})$ level, and their difference can be applied to assess the associated uncertainty. Achieving higher accuracy requires a detailed dynamical model incorporating realistic electromagnetic and strong interactions such as NPHT. 



{\it Summary} --- A comprehensive analysis of pion electroproduction is essential for testing theoretical models and serves as a valuable tool for exploring the internal structure and properties of nucleon excited states. We have extended our previous framework for pion photoproduction to systematically investigate near-threshold pion electroproduction processes. The electric dipole amplitudes in the finite volume show good agreement with recent lattice QCD data. The real and imaginary parts are then extracted with NPHT and brought to the infinite volume. The Lellouch–L{\"u}scher factor can conveniently give the absolute value $|E_{0+}|$ while the factor we find in this letter can provide both the real and imaginary parts of $E_{0+}$ from the finite volume $E_{0+}^L$. In addition, this factor can also be applied to other electroweak amplitudes.

In our approach, pion electroproduction is studied comprehensively with the inclusion of the $\rho$ meson and nucleon excited states, combined with the $\pi N$, $\eta N$, and $K \Lambda$ two-particle channels. We have checked that the $\eta N$ and $K \Lambda$ contributions are small near the threshold. For example, removing their contributions to the physical $E_{0+}$ of the proton target reduces the values by 3.7\% to 4.5\% over the range of $q^2$ values considered. Thus their contributions are small but important in obtaining precision results. For the current lattice results, we find it is sufficient to consider the electromagnetic form factors of nucleons and pions and the $NN\pi$ and $NN\pi\gamma$ vertices to explain the original lattice QCD data. The former are well determined from experiment, and the latter two vertices share only one coupling, $f_{\pi NN}$, because of gauge invariance. In the future, more precise lattice QCD simulations will better constrain these fundamental couplings.

Drawing on our comprehensive formalism, we have also extended the calculation of the electric dipole amplitudes to larger energies, finding that the finite-volume effects on $E_{0+}$ are smaller for the first excited eigenstates than for the ground states. Clearly, further lattice QCD simulations would help us better constrain the electromagnetic properties of the nucleon at higher energies.



{\it Acknowledgments} --- This work is supported by the National Natural Science Foundation of China under Grants No. 12175091, No. 12335001, No. 12247101, the ``111 Center'' under Grant No. B20063, and the innovation project for young science and technology talents of Lanzhou city under Grant No. 2023-QN-107. This research is supported by the University of Adelaide and by the Australian Research Council through Discovery Projects DP210103706 (DBL) and DP230101791 (AWT).

\bibliography{ref}
\end{document}